\documentclass{article}
\usepackage{spconf,amsmath,graphicx}
\usepackage{amsfonts}
\usepackage{graphicx}
\usepackage{float}
\usepackage{subfig}
\usepackage{mathrsfs}
\usepackage{multirow} % 用于合并行
\usepackage{booktabs} % 用于更好的表格线条
\usepackage{tabularx}
\usepackage{cite}

\title{SELD-Mamba: Selective State-Space Model for Sound Event Localization and Detection with Source Distance Estimation}
\name{Da Mu, Zhicheng Zhang*, Haobo Yue, Zehao Wang, Jin Tang, Jianqin Yin\thanks{* Corresponding author.}} 
\address{School of Artificial Intelligence, Beijing University of Posts and Telecommunications, China}

\begin{document}
%\ninept
%
\maketitle
\begin{abstract}
In the Sound Event Localization and Detection (SELD) task, Transformer-based models have demonstrated impressive capabilities. However, the quadratic complexity of the Transformer's self-attention mechanism results in computational inefficiencies. In this paper, we propose a network architecture for SELD called SELD-Mamba, which utilizes Mamba, a selective state-space model. We adopt the Event-Independent Network V2 (EINV2) as the foundational framework and replace its Conformer blocks with bidirectional Mamba blocks to capture a broader range of contextual information while maintaining computational efficiency. Additionally, we implement a two-stage training method, with the first stage focusing on Sound Event Detection (SED) and Direction of Arrival (DoA) estimation losses, and the second stage reintroducing the Source Distance Estimation (SDE) loss. Our experimental results on the 2024 DCASE Challenge Task3 dataset demonstrate the effectiveness of the selective state-space model in SELD and highlight the benefits of the two-stage training approach in enhancing SELD performance.

\end{abstract}
\section{Introduction}
\label{sec:intro}

Sound Event Localization and Detection (SELD) is a multi-task that includes Sound Event Detection (SED) and Direction of Arrival (DoA) estimation. Since its introduction as Task3 of the Detection and Classification of Acoustic Scenes and Events (DCASE) challenge \cite{adavanne2018sound}, SELD has been significantly developed with the use of deep neural network (DNN) models \cite{cao2021improved, hu2022track, niu2023experimental, shul2024cst}, especially those based on Transformer architectures, such as the Event-Independent Network V2 (EINV2) \cite{hu2022track} and CST-former \cite{shul2024cst}. EINV2 employs the Conformer \cite{gulati20_interspeech}, which integrates convolutional layers and multi-head self-attention (MHSA) mechanisms \cite{NIPS2017_3f5ee243} to extract both local and global features. CST-former independently applies attention mechanisms to channel, spectral, and temporal domains. Although Transformer-based models have shown promising results, their quadratic complexity in self-attention renders them computationally inefficient. Furthermore, the 2024 DCASE Challenge Task3 introduces Source Distance Estimation (SDE) for the detected events, which makes the task significantly more challenging.

Utilizing State Space Models (SSMs), which establish long-range context dependencies with linear computational complexity, is expected to overcome the aforementioned limitation. Recently, SSMs, exemplified by Mamba\cite{gu2024mambalineartimesequencemodeling}, have demonstrated their effectiveness across various domains, including natural language processing \cite{yang2024clinicalmambagenerativeclinicallanguage}, computer vision \cite{liu2024vmambavisualstatespace, zhu2024visionmambaefficientvisual}, and speech processing\cite{li2024spmambastatespacemodelneed, jiang2024dualpathmambashortlongterm, zhang2024mambaspeechalternativeselfattention}. However, the design of effective and efficient models using SSMs for SELD has yet to be explored.

% However, the design of effective and efficient models using SSMs for SELD has yet to be explored.

In this paper, we introduce Mamba to SELD, proposing a novel architecture named SELD-Mamba. SELD-Mamba is built upon the robust framework of EINV2, which leverages the Conv-Conformer architecture. Specifically, by replacing the Conformer blocks of EINV2 with bidirectional Mamba (BMamba) blocks, SELD-Mamba aims to enhance the modeling of audio sequence contexts while maintaining linear complexity with sequence length. Furthermore, recognizing the greater challenge of SED and DoA estimation compared to SDE, we employ a two-stage training approach. In the first stage, we focus on the losses for SED and DoA estimation tasks, and in the second stage, we reintroduce the SDE task loss. Our comprehensive experiments on the 2024 DCASE Challenge Task3 dataset highlight the exceptional performance of SELD-Mamba and the effectiveness of the two-stage training method. Compared with EINV2, we achieve superior results by utilizing fewer parameters and reduced computational complexity. In addition to directly improving performance, this work also pioneers the application of SSMs in the field of SELD.
\begin{figure*}[t]
	\centering
	\subfloat[\textnormal{SELD-Mamba Model}]
        {\includegraphics[width=.6\linewidth]{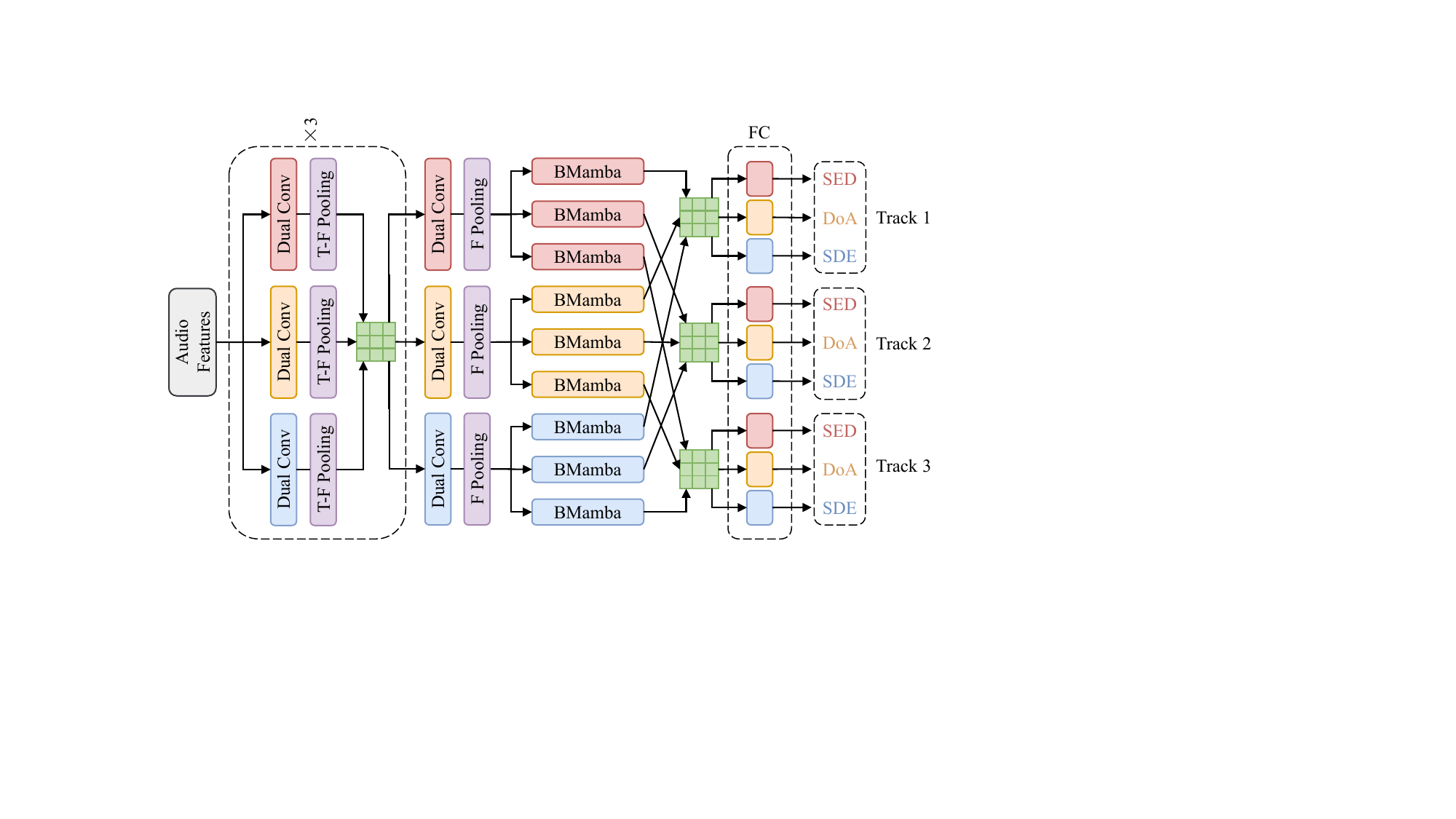} \label{fig1a}}
        \hspace{35pt}
	\subfloat[\textnormal{BMamba Block}]
        {\includegraphics[width=.26\linewidth]{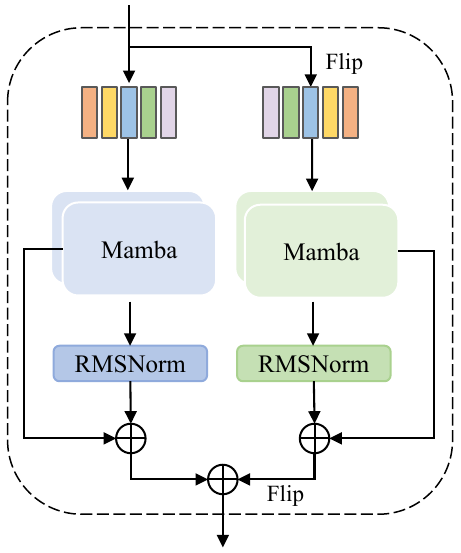} \label{fig1b}}
	\caption{\textnormal{(a) An overview of the proposed SELD-Mamba model, which uses EINV2 as the base model and replaces the Conformer with BMamba. Red, yellow, and blue correspond to SED, DoA estimation, and SDE tasks, respectively. The green boxes signify the soft connections among the three tasks. (b) The description of BMamba block, which handles both forward and backward audio sequences.}}
        \label{SELD-Mamba and BMamba}
\end{figure*}
\section{RELATED WORK: Mamba}
\label{sec:related}
% In the SELD task, the challenge lies in effectively completing the SED, DoA estimation, and SDE tasks simultaneously. This is particularly crucial in scenarios involving overlapping sound events. Early methods introduced the Convolutional Recurrent Neural Network (CRNN) model, which employed CNNs to extract local features followed by RNNs to model long-term dependencies. However, RNNs are difficult to parallelize and struggle with vanishing or exploding gradients. Subsequently, more effective Transformer-based models emerged. For instance, EINV2 uses Conformer, which integrates convolutional layers and multi-head self-attention (MHSA) mechanisms to coherently extract both local and global temporal context information from the feature sequence. CST-former independently applies attention mechanisms to spectral, spatial, and channel domains. However, each method faces its own challenges in computational efficiency and capturing temporal dependencies.

% In addition, to address the issue of overlapping sound events of the same class, researchers proposed class-wise and track-wise output formats. Networks using the class-wise output format merge SED and DoA outputs into a single ACCDOA output, which can be further extended to Multi-ACCDOA format. On the other hand, EINV2, as a representative of the track-wise output format, employs a network with two branches dedicated to the SED and DoA tasks respectively. It utilizes soft connections to enable the network to decide which useful information to exchange and which to retain.
SSM performs a sequence-to-sequence transformation, mapping input $\boldsymbol {x}(t) \in \mathbb{R} $ to output $\boldsymbol{y}(t) \in \mathbb{R} $ through an implicit latent state \( \boldsymbol{h}(t) \in \mathbb{R}^N \), where \( N \) is the dimension of the hidden state, as illustrated in the equation below:
\begin{equation}
\boldsymbol{h'}(t)= \boldsymbol{A}\boldsymbol{h}(t)+\boldsymbol{B}\boldsymbol{x}(t), \quad \boldsymbol{y}(t)=\boldsymbol{C}\boldsymbol{h}(t)
\end{equation}
where \( \boldsymbol{A} \in \mathbb{R}^{N \times N} \), \(\boldsymbol{B} \in \mathbb{R}^{N \times 1} \), and \(\boldsymbol{C} \in \mathbb{R}^{1 \times N} \) represent the state transition matrix, the input projection matrix, and the output projection matrix, respectively. To facilitate the model's application to discrete-time signals, the continuous parameters \(({\Delta}, \boldsymbol{A}, \boldsymbol{B})\) are discretized to their discrete parameters \((\boldsymbol{\Bar{A}}, \boldsymbol{\Bar{B})}\):
\begin{equation}
\boldsymbol{h}_{k}=\boldsymbol{\Bar{A}} \boldsymbol{h}_{k-1}+\boldsymbol{\Bar{B}} \boldsymbol{x}_{k}, \quad \boldsymbol{y}_{k}=\boldsymbol{C} \boldsymbol{h}_{k}
\end{equation}
% A learnable parameter \(\Delta\) balances the focus between the current state and the inputs
To process an input sequence \( \boldsymbol{x} \) of length \( L \) with \( D \) channels, the SSM is applied independently to each channel.

Mamba's innovation lies in its introduction of a selection mechanism within SSMs, achieved by making several parameters \((\Delta, \boldsymbol{B}, \boldsymbol{C})\) functions of the input. This strategy enables Mamba to dynamically focus on or ignore information along the sequence, a capability that is particularly important for effectively detecting overlapping sound events. Additionally, Mamba uses a hardware-aware algorithm, which is able to efficiently compute selective SSMs on modern GPU architectures.

\section{METHOD}
\label{sec:method}
In this section, we will first explain SELD-Mamba, as illustrated in Fig.\ref{SELD-Mamba and BMamba}.\subref{fig1a}, with a focus on the BMamba block. Then, we will introduce the loss function design and outline our two-stage training method.

\subsection{SELD-Mamba}
The SELD-Mamba model utilizes EINV2 as its backbone, a multi-task learning network with two branches dedicated to the SED and DoA estimation tasks. We expand it to three branches by incorporating the SDE task. Additionally, we replace the Conformer blocks with BMamba blocks.

Fig.\ref{SELD-Mamba and BMamba}.\subref{fig1a} illustrates an overview of SELD-Mamba. The model employs CNNs as the encoder and BMamba blocks as the decoder. The final output is produced by fully connected (FC) layers in a track-wise output format, consisting of three tracks. Soft connections are established between the three branches, allowing each to exchange useful information selectively.
\subsubsection{Encoder}
The encoder processes input features extracted from the FOA array signals. Specifically, we extract log-mel spectrogram and Intensity Vectors (IVs), which are then concatenated along the channel dimension, resulting in audio features with a shape of $7\times T\times F$, where $7$ represents channels, $T$ represents the temporal bins, and $F$ represents the frequency bins. The three branches receive different audio features: the SED and SDE branches receive log-mel spectrograms, while the DoA branch receives both log-mel spectrograms and IVs. Each branch contains four Dual Convolutional (Dual Conv) layers. Time-Frequency (T-F) pooling layers are applied after the first three Dual Conv layers, while only F pooling is applied after the final Dual Conv layer. This results in a tensor with a shape of $512\times T/8 \times F/16$. This tensor is then reshaped and applied frequency average pooling, producing $T/8\times 512$ dimensional feature embedding. The $T/8$ dimension ensures alignment with the temporal resolution of the target label.

In addition, we employ cross-stitch \cite{misra2016cross} as soft connections to facilitate the exchange of useful information between each branch, represented as follows:
\begin{equation}
    \left[\hat{\boldsymbol{x}}^{\mathrm{SED}}, \hat{\boldsymbol{x}}^{\mathrm{DoA}}, \hat{\boldsymbol{x}}^{\mathrm{SDE}}\right]^{\top}=\boldsymbol{\alpha}\left[\boldsymbol{x}^{\mathrm{SED}}, \boldsymbol{x}^{\mathrm{DoA}},{\boldsymbol{x}}^{\mathrm{SDE}}\right]^{\top}
\end{equation}
where $\hat{\boldsymbol{x}}^{\mathrm{SED}}$, $\hat{\boldsymbol{x}}^{\mathrm{DoA}}$, and $\hat{\boldsymbol{x}}^{\mathrm{SDE}}$ are the new features, and $\boldsymbol{x}^{\mathrm{SED}}$, $\boldsymbol{x}^{\mathrm{DoA}}$, and $\boldsymbol{x}^{\mathrm{SDE}}$ are the original features. $\boldsymbol{\alpha}$ is a 3 × 3 matrix that denotes learnable parameters.

\begin{figure}[htbp]
	\centering
	{\includegraphics[width=.55\linewidth]{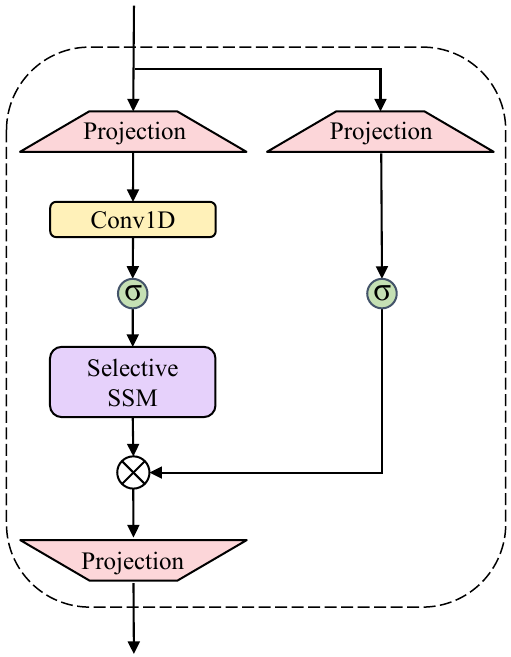} \label{fig2a}}
	\caption{\textnormal{The illustration of Mamba layer. $\sigma $ denotes the SiLU activation.}} 
    \label{Mamba}
\end{figure}

\subsubsection{Decoder}
As the decoder, we replace the Conformer with BMamba. Each branch utilizes three parallel BMamba blocks, corresponding to the three tracks of the output. The Mamba architecture is limited to capturing only historical information about the input due to its causal processing. To leverage future context, we borrow the BMamba design from \cite{li2024spmambastatespacemodelneed}. This design processes the original and flipped input sequences through two separate Mamba components, as shown in Fig.\ref{SELD-Mamba and BMamba}.\subref{fig1b}. A Mamba component is composed of two Mamba layers, with the structure of one Mamba layer illustrated in Fig.\ref{Mamba}.

Taking forward audio sequence as an example, we begin with an input $\boldsymbol{u} \in \mathbb{R}^{L \times D}$, where $L$ is the number of frames and $D$ matches the encoder dimension. A linear layer projects $\boldsymbol{u}$ to $\boldsymbol{\hat{u}} \in \mathbb{R}^{L \times E}$, where $E = 2D$, representing the dimension expanded by a factor of 2. Another linear layer projects $\boldsymbol{u}$ to $\boldsymbol{z} \in \mathbb{R}^{L \times E}$, which will be used to gate the outputs of SSM:
\begin{equation}
    \boldsymbol{\hat{u}} = \mathrm{Linear_{input}}(\boldsymbol{u}), \quad \boldsymbol{z} = \mathrm{Linear_{gated}}(\boldsymbol{u})
\end{equation}
Next, $\boldsymbol{\hat{u}}$ is processed through convolution and SiLU activation, resulting in $\boldsymbol{x}$: 
\begin{equation}
    \boldsymbol{x} = \sigma (\mathrm{Conv1D}(\boldsymbol{\hat{u}})) 
\end{equation}
where $\sigma$ represents the SiLU function. Then, $\boldsymbol{x}$ serves as the input to the SSM, as described in Section \ref{sec:related}. The outputs of the SSM are gated by $\sigma(\boldsymbol{z})$:
\begin{equation}
    \boldsymbol{y} = \sigma (\boldsymbol{z}) \otimes \mathrm{SSM}(\boldsymbol{x})
\end{equation}
A linear projection is then applied to obtain the final output:
\begin{equation}
    \boldsymbol{\hat{y}} = \mathrm{Linear_{output}}(\boldsymbol{y}) 
\end{equation}
$\boldsymbol{\hat{y}}$ is used as the input for the next Mamba layer.

We employ RMSNorm \cite{zhang2019root} to normalize the outputs of the Mamba layers. The outputs obtained from the backward Mamba are then reversed to the forward direction and fused with the outputs from the forward Mamba through element-wise addition.

\subsection{Loss Function}
\label{sec:loss fuction}
For the loss function, we utilize frame-level Permutation Invariant Training (PIT) \cite{cao2021improved} to compute the total loss:
\begin{equation}
\label{eq:PIT}
\begin{split}
&\mathcal{L}_{PIT}(o)= \\
&\min_{\alpha \in \mathbf{P}(o)} \sum_{M}\left\{\lambda_1 {\mathcal{L}_{SED}}(o)+\lambda_2 {\mathcal{L}_{DoA}}(o)+\lambda_3 {\mathcal{L}_{SDE}}(o) \right\}
\end{split}
\end{equation}  
Where $\alpha \in P(o)$ denotes one of the possible permutations. $\mathcal{L}_{SED}$ is binary cross entropy loss for SED, $\mathcal{L}_{DoA}$ is mean squared error loss for DoA, and $\mathcal{L}_{SDE}$ is L1 loss for SDE. $\lambda_1$, $\lambda_2$, and $\lambda_3$ are weights for the SED, DoA, and SDE losses, respectively. The permutation yielding the minimum loss is selected for optimization.

In comparison to the SDE task, the SED and DoA estimation tasks are considerably more challenging. Therefore, we introduce a two-stage training method for SELD-Mamba. Initially, we focus on optimizing the SED and DoA losses by setting $\lambda_3$ to 0 and assigning weights of $\lambda_1 = 25$ and $\lambda_2 = 5$. In the second stage, we reintroduce the SDE loss by adjusting $\lambda_3$ to 3. This two-stage training approach is essential for achieving balanced performance across the three tasks.

\begin{table*}[ht]
\belowrulesep=0pt
\aboverulesep=0pt
\centering
\caption{Performance comparison of SELD-Mamba with other models on the dev-test dataset. The MACs were calculated by processing a 1-second audio sequence on GPU.}
\label{table 1}
\begin{tabular}{cccccccc}
\toprule
Model & Training & Params(M) & Macs (G/s) & ${F}_{20^\circ}$↑
 & ${DOAE}$↓ & ${RDE}$↓ & ${SELD}_{\mathrm{score}}$↓ \\
\midrule
2024 Baseline \cite{krause2024soundeventdetectionlocalization}  & -  & \textbf{0.74} & \textbf{0.03}& 13.1 & 36.9 & 33.0 &0.468 \\
EINV2 \cite{hu2022track} & Unified-Training & 127.93 & 34.36 & 26.8& 28.7 &32.9 & 0.407  \\
\cline{1-8}
\multirow{3}{*}{SELD-Mamba (ours)}
& Unified-Training & \multirow{3}{*}{75.14} & \multirow{3}{*}{6.35} & 26.2& 27.3& 28.6& 0.392 \\
 & Stage-1& & & \textbf{27.3} & \textbf{24.9}& 62.6 &0.497 \\
& Stage-2& & & \textbf{27.3} &25.1 &\textbf{27.8} &\textbf{0.381} \\
\bottomrule
\end{tabular}
\end{table*}

\section{EXPERIMENTS}
\label{sec:experiments}
\subsection{Implementation Details}
The proposed method was evaluated using the official development \cite{Shimada2023starss23_arxiv} and synthetic dataset \cite{krause_2024_10932241} of the 2024 DCASE Challenge Task3, without employing data augmentation. The model was only trained on FoA array signals. Audio clips were segmented into non-overlapping 5-second fixed segments, with a sampling rate of 24 kHz. A Short Time Fourier Transform (STFT) was applied using a 1024-point Hanning window and a hop size of 300. Subsequently, log-mel spectrograms and IVs were generated in the log-mel space with 128 frequency bins. The corresponding audio features were fed into their respective branches. The output includes three tracks, enabling the detection of up to three overlapping sound events. The AdamW \cite{loshchilovdecoupled} optimizer was employed for training over 80 epochs. The initial learning rate was set at 0.0003 and halved after 65 epochs. We employed two training methods: unified-training and two-stage training. For unified-training, we set $\lambda_1 = 25$, $\lambda_2 = 5$, and $\lambda_3 = 1$. The details of the two-stage training method are provided in Section \ref{sec:loss fuction}

For evaluation, we used the location-dependent F-score (${F}_{20^\circ}$), class-dependent DoA error (${DOAE}$), and class-dependent relative distance error (${RDE}$). To compare model performance comprehensively, we introduced the ${SELD}_{\mathrm{score}}$, the average of the three metrics. We also reported the number of parameters and computational cost of different models.

\subsection{Performance Comparison}

To validate the proposed model, we compare SELD-Mamba with the 2024 Baseline \cite{krause2024soundeventdetectionlocalization} and EINV2 \cite{hu2022track} models. 2024 Baseline employs a convolutional recurrent neural network (CRNN) with two additional MHSA layers. EINV2 uses Conv-Conformer architecture. The comparison of model performance is presented in Table \ref{table 1}.

Using the unified-training approach, SELD-Mamba outperforms the 2024 Baseline across all metrics. Compared to EINV2, our ${F}_{20^\circ}$ slightly lags behind, but our ${DOAE}$, ${RDE}$, and ${SELD}_{\mathrm{score}}$ are superior. Notably, SELD-Mamba achieves these results with significantly fewer parameters and lower computational complexity. This underscores the effectiveness and efficiency of SELD-Mamba in handling the SELD task.

When utilizing the two-stage training approach, our model attains the best ${F}_{20^\circ}$ and ${DOAE}$ in the first stage. Interestingly, even with the SDE loss weight set to 0, ${RDE}$ achieved 62.6. This may be attributed to the model learning distance information from the DoA estimation task. Upon incorporating the SDE loss in the second stage, ${SELD}_{\mathrm{score}}$ achieves 0.381. This demonstrates the effectiveness of the two-stage training approach in balancing results across different tasks and enhancing performance. 

\subsection{Ablations}
\subsubsection{Input features of SDE branch}

\begin{table}[ht]
\belowrulesep=0pt
\aboverulesep=0pt
\centering
\caption{Performance of SELD-Mamba with different input features for the SDE branch.}
\label{table 2}
\begin{tabular}{c|cccc}
\toprule
Features & ${F}_{20^\circ}$↑
 & ${DOAE}$↓ & ${RDE}$↓ & ${SELD}_{\mathrm{score}}$↓  \\
\midrule
log-mel & \textbf{26.2} & 27.3 & \textbf{28.6} & \textbf{0.392} \\
+ IVs & 24.3& \textbf{26.0} &34.5 & 0.416  \\
\bottomrule
\end{tabular}
\end{table}

\begin{table}[ht]
\belowrulesep=0pt
\aboverulesep=0pt
\centering
\caption{Performance of SELD-Mamba with different $\lambda_3$ values in the second stage.}
\label{table 3}
\begin{tabular}{c|cccc}
\toprule
$\lambda_3$ & ${F}_{20^\circ}$↑
 & ${DOAE}$↓ & ${RDE}$↓ & ${SELD}_{\mathrm{score}}$↓ \\
\midrule
1 & \textbf{29.8} & 24.5 & 33.7 & 0.392 \\
2 & 28.0& 24.7 &31.9 & 0.392  \\
3 & 27.3 & 25.1 & \textbf{27.8} & \textbf{0.381} \\
4 & 28.2 & 24.2 & 30.2 &0.385 \\
5 & 27.8 & \textbf{24.0} & 29.4 &0.383 \\
\bottomrule
\end{tabular}
\end{table}

To find the best input features for the SDE branch, we tested two types of features, with the results shown in Table \ref{table 2}. Compared to using log-mel spectrograms alone, adding IVs only improved the $DOAE$. This might be due to IVs providing more source direction information, but not necessarily offering additional benefits for sound class perception and source distance estimation. Therefore, we chose to use only log-mel spectrograms as the input for the SDE branch.

\subsubsection{Loss weight of SDE task in the second stage}

The loss weight of the SDE task in the second stage affects the performance balance across different tasks. We adjusted the value of $\lambda_3$, and the results are presented in Table \ref{table 3}. The results indicate that $\lambda_3 = 3$ achieves a balanced performance and results in the best $SELD_{\mathrm{score}}$.

\section{CONCLUSION}
\label{sec:conclusion}

In this paper, we introduce SELD-Mamba, a novel SELD architecture. By integrating the BMamba module into EINV2, SELD-Mamba is able to capture long-range contextual information while maintaining computational efficiency. Additionally, we employ a two-stage training approach to balance performance across different tasks. Our experimental results demonstrate the superior performance of SELD-Mamba and validate the effectiveness of the selective state-space model in the SELD task.

\bibliographystyle{IEEEtran}
\bibliography{refs}

\end{document}